\documentclass[aps,prl,reprint]{revtex4-1}
\usepackage{blindtext}
\usepackage{hyperref}
\usepackage{hypernat}

\usepackage{enumerate}
\usepackage{amsmath,amssymb,epsfig,cancel}
\usepackage{color}
\usepackage{subfig}
\usepackage{graphicx}
\usepackage{caption}
\usepackage{braket}
\usepackage[english]{babel}

\renewcommand{\d}[1]{\ensuremath{\operatorname{d}\!{#1}}}

\usepackage{physics}
\usepackage{cancel}

\begin{document}
\title{Non-integrability of the $\Omega$-deformation}
\author{Kostas Filippas}
\email{kphilippas@hotmail.com}
\affiliation{Department of Physics, Swansea University, Swansea SA2 8PP, United Kingdom}

\begin{abstract}
We study integrability on the supergravity vacuum dual to the field theoretical $\Omega$-deformation of $\mathcal{N}=4$ super Yang-Mills theory. The deformation manifests itself as turning on a Kalb-Ramond field on the (Euclidean) AdS$_5\times$S$^5$, while the associated $H_3$ flux ignores half of the geometric isometries. By constructing appropriate string embeddings that incorporate the essential $H_3$ flux contribution on this background, we study their fluctuations through the associated Hamiltonian systems. Each and every case demonstrates that the string exhibits non-integrable dynamics, which in turn suggests that the $\Omega$-deformation does not preserve classical integrability.
\end{abstract}

\maketitle

\section{Introduction}
Integrability possesses a prominent role in field theory, providing a rich variety of conserved quantities and, at the same time, informing us of the theory being solvable for all values of its coupling constant, \cite{Beisert:2010jr,Torrielli:2016ufi,Zarembo:2017muf}. Since holography relates the worldsheet of the superstring to a quantum field theory, spotting integrable field theories has largely become a matter of studying the various backgrounds in string theory. Nevertheless, integrable structures are relatively rare and hard to find. This is due to integrability relying on the existence of a Lax connection on the cotangent bundle of the theory, while no systematic way of building such a connection is available to date. In fact, there is not even an apparent reason to decide whether a Lax connection exists or not in the first place, except if we are already aware of the theory being non-integrable. In this sense, integrable theories are mainly obtained by structure-preserving deformations of well-known integrable models, \cite{Lunin:2005jy,Sfetsos:2013wia,Delduc:2014kha,Borsato:2016pas}.\vspace{0.09cm}

Through the subjective constraints of the methods of integrability, analytic non-integrability emerges in a dialectic way. Considering worldsheet embeddings in string theory, differential Galois theory through Kovacic's theorem \cite{Kovacic} acts on their associated Hamiltonian systems producing a statement about the (non-) integrability of their structure, \cite{Basu:2011fw,Chervonyi:2013eja,Stepanchuk:2012xi,Giataganas:2013dha,Giataganas:2014hma,Giataganas:2017guj,Roychowdhury:2017vdo,Roychowdhury:2019olt,Nunez:2018ags,Nunez:2018qcj,Filippas:2019puw,Filippas:2019ihy}. Since an integrable theory exhibits its homonymous property universally, even a single non-integrable sector of the associated supergravity background $-$ corresponding to a particular string embedding $-$ is enough to declare the whole theory as non-integrable.\vspace{0.09cm}

On another approach \cite{Wulff:2017lxh,Wulff:2017vhv,Wulff:2017hzy,Wulff:2019tzh}, S-matrix factorization on the worldsheet theory produces certain conditions of non-integrability, while quite recently a reconciliation began to arise between both non-integrability tools, \cite{Giataganas:2019xdj}.\vspace{0.08cm}

A particular supergravity background which deserves the attention of our non-integrability methods was recently discovered in \cite{Bobev:2019ylk}. Neglecting an unimportant warp factor, this background is the holographic dual of the four-dimensional, $\Omega$-deformed $\mathcal{N}=4$ super Yang-Mills (SYM) theory. In the same vein of the supergravity realization of the $\Omega$-deformation, a similar study was also recently performed in \cite{BenettiGenolini:2019wxg}. $\Omega$-deformation was originally introduced in  \cite{Nekrasov:2002qd} as a method of calculating the path integral of four-dimensional $\mathcal{N}=2$ gauge theories, through supersymmetric localization. Since then, the deformation and its associated Nekrasov partition function, have produced numerous exact results on supersymmetric quantum field theories on curved manifolds, as well as having laid the foundations for both the Nekrasov-Shatashvili \cite{Nekrasov:2009rc} and the AGT \cite{Alday:2009aq} correspondences. The background we consider is a deformation of AdS$_5\times$S$^5$ in type IIB theory that preserves 16 supercharges, while the $\Omega$-deformation manifests itself in this dual gravity as turning on a Kalb-Ramond field (and a $C_2$ RR form). Interestingly, the associated $H_3$ flux inter-binds the whole geometry and breaks part of the bosonic symmetries of AdS$_5\times$S$^5$, both being facts that make this background intractable to classic integrability methods.\vspace{0.05cm}

The study of non-integrability on this particular background is of interest, since there are significant suggestions linking integrable structures and the $\Omega$-deformation in the present literature. In particular, a connection has been established between the $\Omega$-deformed $\mathcal{N}=2$ gauge theory and quantum integrable Hamiltonian systems, see \cite{Nekrasov:2009rc,Alday:2009aq,Nekrasov:2009ui,Nekrasov:2010ka} or the more recent \cite{Fioravanti:2019vxi}. Similar work has been done, \cite{Orlando:2018kms,Sekiguchi:2019nhn}, on a string theory realization \cite{Hellerman:2011mv,Reffert:2011dp,Hellerman:2012zf} of the $\Omega$-deformation, where the resulting models were associated with the TsT subclass of the Yang-Baxter deformation. Considering all these integrable aspects of the $\Omega$-deformation, an indication of non-integrability would consequently suggest an interesting antithesis, worthy of further study.\vspace{0.04cm}

In this letter, after a complete symmetry analysis on the ten-dimensional $\Omega$-deformed supergravity background, we accordingly construct string embeddings that are dynamical on the asymmetric directions. We do so, in order to have a better chance to spot non-integrable behavior. We then find simple solutions on the equations of motion and let the string fluctuate around them, along each dimension. As it turns out, in each case, one of the fluctuations exhibits a non-Liouvillian solution in terms of the Bessel function of the first kind, yielding the classical non-integrability of our embedding and, therefore, of the whole background under consideration.

\section{The supergravity solution}
The supergravity background dual to the $\Omega$-deformation of $\mathcal{N}=4$ SYM at the conformal point was introduced in \cite{Bobev:2019ylk}. Neglecting the existence of a warp factor \footnote{This is equivalent to setting $w=0$ in \cite{Bobev:2019ylk}. This parametrizes a VEV of a scalar field in a representation of the SO(6) of $\mathcal{N}=4$ SYM. In the dual gravity, it generates a distribution of smeared D3-branes.} which we can set, along with the radii of the space, to unity we obtain the vacuum

\begin{equation}
\begin{split}
\d s^2&=\frac{\d{\vec{x}_4^2}+\d z^2}{z^2}+\d\theta^2-\sin^2\theta\d\phi^2+\cos^2\theta\d\Omega^2_3\\
B_2&=ig_sC_2=-\frac{\beta e^{-\phi}\sin\theta}{4z}\left(\d x_1\wedge\d x_2+\d x_3\wedge\d x_4\right)\\
F_5&=-\frac{i}{g_s}(1+\star_{10})\,\dd\left(\frac{1}{z^4}\right)\wedge\mbox{vol}_4\hspace{1cm}e^\Phi=g_s
\end{split}\label{PoincareOmega}
\end{equation}
where $g_s$ is the string coupling and vol$_4$ the volume of the $\mathbb{R}^4$ subspace. $\beta\in\mathbb{R}^+$ is the deformation parameter in the dual field theory, which was identified with the linear combination $\epsilon_1+\epsilon_2$ in \cite{Hama:2012bg,Klare:2013dka}. Thus, the $\Omega$-deformation manifests itself as turning on a Kalb-Ramond field (and a $C_2$ RR field) on the integrable H$_5$ $\times$ dS$_5$.\vspace{0.15cm}

Since the internal space of the IIB background (\ref{PoincareOmega}) is a deformation of the five-dimensional de Sitter space, this implies that the background is actually a solution of type IIB* supergravity \cite{Hull:1998vg,Hull:1998ym}. Continuing as $\phi\rightarrow i\varphi$, we obtain the Euclidean AdS$_5\times$S$^5$. The vacuum preserves 16 supercharges and it is the supergravity dual of $\mathcal{N}=4$ SYM. Interestingly, the non-trivial $H_3$ flux inter-binds the geometric subspaces and breaks part of the bosonic symmetries of AdS$_5\times$S$^5$, both facts that make the background intractable to classic integrability methods.\vspace{0.15cm}

While the geometry in (\ref{PoincareOmega}) looks like a peculiar continuation of AdS$_5\times$S$^5$, on which the string dynamics could be qualitatively questioned, it is not quite unfamiliar. In fact, it was obtained in \cite{Dobashi:2002ar} by a double Wick rotation on AdS$_5\times$S$^5$ (in our notation w.r.t to the $\mathbb{R}^4$ time $t\equiv x_1$ and $\phi$), as a natural formulation on which the holographic principle $-$ for the Penrose limit $-$ naturally associates the bulk with the boundary. In particular, it was shown that, for the BMN string on this geometry, the bulk-to-boundary trajectories are interpreted as a tunneling phenomenon and thus that the BMN boundary-to-boundary correlations are holographically well defined.

\section{symmetries of the background}
Since the $\Omega$-deformation is realized on the background as a $B_2$-field $-$ that obviously does not respect part of the geometric isometries $-$ it is instructive to perform a symmetry analysis on its associated $H_3$ flux. Noting that the geometry (\ref{PoincareOmega}) is a product space and thus its Killing vectors (KVs) are decoupled for the two subspaces, we shall vary $H_3$ separately along $H^5$ and dS$_5$.

Hence, if $K$ is a KV on H$_5$, then the vanishing of the Lie derivative $\mathcal{L}_KH_3=0$ is solved for the vectors
\begin{equation}
\begin{split}
K_{R_{12}}&=x_1\partial_2-x_2\partial_1\\
K_{R_{34}}&=x_3\partial_4-x_4\partial_3\\
K_{SC_i}&=\partial_i\hspace{2cm}i=1,...,4
\end{split}\label{H5isometriesH3}
\end{equation}
namely two SO(4) rotations on $\mathbb{R}^4$ and the four SO(1,1) special conformal Killing vectors \footnote{$K=\partial_i, i=1,...,4$, are translations on $\mathbb{R}^4$ and  special conformal transformations on H$^5$.} (SCKVs) on H$_5$. As far as the KVs of dS$_5$ are concerned, the only non trivial KV that leaves $H_3$ invariant is
\begin{equation}
K_B=e^{-\phi}\left(\cot\theta\:\cos\alpha\:\partial_\phi\:+\:\cos\omega_1\:\partial_\theta\:+\:\tan\theta\:\sin\omega_1\:\partial_{\omega_1}\right)\label{dS5isometriesH3}
\end{equation}
where $\omega_1$ is an angle in $\Omega_3$ of dS$_5$ \footnote{There are two more, one for each angle $\omega_2,\omega_3$ in the 3-sphere, $\dd\Omega_3^2=\dd\omega_1^2+\sin^2\omega_1\dd\Omega_2^2$, of dS$_5$. It doesn't make any difference in this problem.}. This rotation is identified as an SO(1,1) boost of the SO(1,5) isometry. The rest of the KVs of dS$_5$ that preserve $H_3$ are trivial, namely the six SO(4) rotations of $\Omega_3$ inside dS$_5$.\vspace{0.15cm}

Note that the symmetry analysis on the background (\ref{PoincareOmega}) is of twofold interest. First, it reveals the action of the $\Omega$-deformation on the symmetry structure of the dual supergravity. Most importantly for our non-integrability method, though, it serves as a beacon on how to push our bosonic string towards a less symmetric embedding, the latter having a better chance to exhibit non-integrable dynamics.

\section{String dynamics}
\paragraph{\textbf{The first embedding}} The bosonic string dynamics emerges from the non-linear $\sigma$-model, in conformal gauge,
\begin{equation}
S_P\:=\:\frac{1}{4\pi\alpha'}\int_\Sigma\dd^2\sigma\,\partial_aX^\mu\partial_bX^\nu\left(g_{\mu\nu}\eta^{ab}+B_{\mu\nu}\epsilon^{ab}\right)\label{PolyakovAction}
\end{equation}
where the string coordinates' $X^\mu(\tau,\sigma)$ equation of motion is supplemented by the Virasoro constraint $T_{ab}=0$, where the worldsheet energy-momentum tensor is given by
\begin{equation}
T_{ab}\:=\:\frac{1}{\alpha'}\left(\partial_aX^\mu\partial_bX^\nu g_{\mu\nu}-\frac{1}{2}\eta_{ab}\eta^{cd}\partial_cX^\mu\partial_dX^\nu g_{\mu\nu}\right)\label{EMtensor}
\end{equation}
with $\tau, \sigma$ being the worldsheet coordinates. Having differential Galois theory in mind, we desire a string embedding that produces second order, \textit{ordinary} linear differential equations of motion. This means that the string coordinates must be $X^\mu=X^\mu(\tau)$ or $X^\mu=X^\mu(\sigma)$. For a closed string in type II theory, this translates into wrapping the string around compact coordinates.\vspace{0.15cm}

Since H$_5\times$dS$_5$ is integrable, our chance to spot non-integrable behavior lies along the $H_3$ flux. Hence, most importantly, our embedding should incorporate dynamics along the $H_3$ flow. The $B_2$ field component(s) $B_{x_1x_2}$ (and $B_{x_3x_4}$) is non-vanishing on the $\sigma$-model (\ref{PolyakovAction}) only for the choice $-$ in these coordinates $-$ $x_1=x_1(\tau)$ and $x_2=x_2(\sigma)$, or vice versa. However, such a $\sigma$-dependence produces partial differential equations of motion for a closed string and, thus, it must be excluded.\vspace{0.15cm}

The resolution comes by changing our coordinates on the $\mathbb{R}^4$ subspace of $H_5$, from Cartesian to spherical, as\vspace{0.1cm}
\begin{equation}
\d{\vec{x}_4^2}=d r^2+r^2\left(\d\psi^2+\sin^2\psi\d\chi^2+\sin^2\psi\sin^2\chi\d{\xi}^2\right)
\end{equation}
with the old coordinates depending on the new ones as
\begin{equation}
\begin{split}
x_1&=r\cos\psi\\\
x_2&=r\sin\psi\cos\chi\\
x_3&=r\sin\psi\sin\chi\cos\xi\\
x_4&=r\sin\psi\sin\chi\sin\xi
\end{split}\label{CartToSphere}
\end{equation}
In this $\mathbb{R}^4$ subspace, we can choose the embedding $r=r(\tau)$, $\chi=\chi(\tau)$, $\xi=\kappa\sigma$, and $\psi=\pi/2$. Since $H_3$ is invariant under only two out of the six SO(4) rotations of $\mathbb{R}^4$, we set $\psi=\pi/2$ but we leave $\chi=\chi(\tau)$ in order to have some portion of $\mathbb{R}^4$ rotations that can bring the equations of motion to the test. The same symmetry analysis also showed that $z$ is non-trivially involved in $H_3$ and thus we let $z=z(\tau)$.\vspace{0.15cm}

As far as dS$_5$ is concerned, we choose $\theta=\theta(\tau)$ and $\phi=\phi(\tau)$ which also parametrize $H_3$ non-trivially. The $\Omega_3$ of dS$_5$ with line element
\begin{equation}
\d\Omega_3^2=\d\omega_1^2+\sin^2\omega_1\d\omega_2^2+\sin^2\omega_1\sin^2\omega_2\d\omega_3^2
\end{equation}
is not involved in the $H_3$ flux, the latter being invariant under its SO(4) rotations, and thus we set $\omega_1=\omega_2=\pi/2$, while we wrap the string as $\omega_3=\nu\sigma$ to reinforce the stringy character of the embedding. Indeed, both wrappings $-$ along $\xi$ and $\omega_3$ $-$ turn out to play a crucial role in surfacing the full power of the $H_3$ dynamical contribution. Also, notice that having non-dynamical $\omega_i$ prevents the string soliton from boosting symmetrically as in (\ref{dS5isometriesH3}). Overall, the string embedding reads
\begin{equation}
\begin{split}
r=r(\tau)\hspace{0.6cm}\chi=\chi(\tau)\hspace{0.6cm}\psi=\frac{\pi}{2}\hspace{0.6cm}\xi=\kappa\sigma\hspace{0.6cm}z=z(\tau)\hspace{0.3cm}\\[10pt]
\theta=\theta(\tau)\hspace{0.7cm}\phi=\phi(\tau)\hspace{0.7cm}\omega_1=\omega_2=\frac{\pi}{2}\hspace{0.7cm}\omega_3=\nu\sigma\hspace{0.5cm}
\end{split}\label{stringSoliton}
\end{equation}
where $\kappa,\nu\in\mathbb{Z}$. Translating the $B_2$ field according to the map (\ref{CartToSphere}) and the above embedding as
\begin{equation}
B_2=-\frac{\beta e^{-\phi}\sin\theta}{4}\left(r\sin^2\chi\d r\wedge\d\xi+r^2\sin\chi\cos\chi\d\chi\wedge\d\xi\right)
\end{equation}
then the $\sigma$-model (\ref{PolyakovAction}) on the embedding (\ref{stringSoliton}) reduces into the Lagrangian density
\begin{equation}
\begin{split}
\mathcal{L}\;&=\;\frac{-\dot{r}^2-r^2\dot{\chi}^2+\kappa^2r^2\sin^2\chi-\dot{z}^2}{z^2}-\dot{\theta}^2+\sin^2\theta\:\dot{\phi}^2\\
&+\nu^2\cos^2\theta-\frac{\beta\kappa\,e^{-\phi}\sin\theta}{2z}\left(r\sin^2\chi\:\dot{r}+r^2\cos\chi\sin\chi\:\dot{\chi}\right)
\end{split}\label{Lagrangian}
\end{equation}
where the dot implies derivation w.r.t to the worldsheet time $\tau$. For our particular string embedding, the equations of motion for this Lagrangian are equivalent to those of the $\sigma$-model and read
\begin{equation}
\begin{split}
4\ddot{r}\:=\:\beta\kappa\,e^{-\phi}\,r\sin^2\chi&\left(z\sin\theta\:\dot{\phi}+\sin\theta\:\dot{z}-z\cos\theta\:\dot{\theta}\right)\\
-&4\,r\left(\kappa^2\sin^2\chi-\dot{\chi}^2\right)+\frac{8\dot{r}\dot{z}}{z}\label{EOMr}
\end{split}
\end{equation}\vspace{-15pt}
\begin{equation}
\begin{split}
4r\ddot{\chi}\:=\:\beta\kappa\,e^{-\phi}\,r&\cos\chi\sin\chi\left(z\sin\theta\:\dot{\phi}+\sin\theta\:\dot{z}-z\cos\theta\:\dot{\theta}\right)\\
&-4\,r\kappa^2\cos\chi\sin\chi+\frac{8r\dot{\chi}\dot{z}}{z}-8\dot{r}\dot{\chi}\label{EOMchi}
\end{split}
\end{equation}\vspace{-10pt}
\begin{equation}
\begin{split}
4z\ddot{z}\:=\:-\beta\kappa\,e^{-\phi}\,&rz\sin\theta\sin\chi\Big(\sin\chi\:\dot{r}+r\cos\chi\:\dot{\chi}\Big)\\
+&4\,r^2\left(\kappa^2\sin^2\chi-\dot{\chi}^2\right)+4\left(\dot{z}^2-\dot{r}^2\right)
\end{split}\label{EOMz}
\end{equation}
\begin{equation}
\begin{split}
4\ddot{\theta}\:=\:2\nu^2\sin2\theta\:+\:&\beta\kappa\,e^{-\phi}\,r\cos\theta\sin\chi(\sin\chi\:\dot{r}+r\cos\chi\:\dot{\chi})\\
&\hspace{3.5cm}-2\sin2\theta\:\dot{\phi}^2\label{EOMtheta}
\end{split}
\end{equation}\vspace{-10pt}
\begin{equation}
4\sin\theta\:\ddot{\phi}\:=\:-8\cos\theta\,\:\dot{\theta}\dot{\phi}\,+\,\beta\kappa \,e^{-\phi}\:r\sin\chi(\sin\chi\:\dot{r}+r\cos\chi\:\dot{\chi})\label{EOMphi}
\end{equation}\\
These equations are constrained by the worldsheet equation of motion, i.e. the Virasoro constraint
\begin{equation}
\begin{split}
2\,T_{\tau\tau}=\:2\,T_{\sigma\sigma}\:=\:\frac{\dot{r}^2+r^2\dot{\chi}^2+\kappa^2r^2\sin^2\chi+\dot{z}^2}{z^2}-\sin^2\theta\:&\dot{\phi}^2\\
+\dot{\theta}^2+\nu^2\cos^2\theta\:=\:&0\\
T_{\tau\sigma}=0\hspace{7cm}
\end{split}\label{VC}
\end{equation}
The worldsheet energy-momentum tensor is conserved, $\nabla_aT^{ab}=0$, since $\partial_\tau T_{\tau\tau}=\partial_\sigma T_{\sigma\sigma}=0$ on the equations of motion (\ref{EOMr})-(\ref{EOMphi}). This compliance of the worldsheet constraints with the string coordinates' equations of motion yield, also, the consistency of our embedding.\vspace{0.15cm}

Transforming into the Hamiltonian formulation, our worldsheet theory reduces to a simple particle system with conjugate momenta
\begin{equation}
\begin{split}
p_r&=-\frac{2\dot{r}}{z^2}-\frac{\beta\kappa\,e^{-\phi}\sin\theta}{2z}r\sin^2\chi\,,\hspace{0.4cm}p_z=-\frac{2\dot{z}}{z^2}\,,\hspace{0.3cm}p_\theta=-2\dot{\theta}\\
p_\chi&=-\frac{2r^2\dot{\chi}}{z^2}-\frac{\beta\kappa\,e^{-\phi}\sin\theta}{2z}r^2\sin\chi\cos\chi\,,\hspace{0.6cm}p_\phi=2\sin^2\theta\dot{\phi}
\end{split}\label{conjMomenta}
\end{equation}
and Hamiltonian density
\begin{equation}
\begin{split}
\mathcal{H}\:&=\:-\frac{z^2}{4r^2}\left(p_\chi+\frac{\beta\kappa\,e^{-\phi}\sin\theta\,r^2\sin\chi\cos\chi}{2z}\right)^2-\frac{z^2p_z^2}{4}\\
&-\frac{z^2}{4}\left(p_r+\frac{\beta\kappa\,e^{-\phi}\sin\theta\,r\sin^2\chi}{2z}\right)^2+\frac{p_\phi^2}{4\sin^2\theta}-\frac{p_\theta^2}{4}\\[5pt]
&\hspace{4cm}-\kappa^2r^2\sin^2\chi-\nu^2\cos^2\theta
\end{split}\label{Hamiltonian}
\end{equation}
Of course, Hamilton's equations of motion on the above system coincide with the Euler-Lagrange equations (\ref{EOMr})-(\ref{EOMphi}). In this effective particle system, the masses are determined by the geometry and they can be read off through the kinetic terms. The string winding modes manifest themselves as a non-trivial potential on the particle dynamics, while the $\Omega$-deformation (i.e. the $H_3$ flux) is realized as a magnetic disturbance on the particle kinematics.\vspace{0.15cm}

Before we proceed to analyze the dynamics, a crucial comment is in place. Usually, in this kind of Hamiltonian analysis on a string embedding we have a well defined equation of motion for the target space time, which always gives the energy of the string as its first integral and so on. Although not often emphasized, this is essential for a string state to be holographically associated with a dual operator, even if we don't know what that operator looks like. And we do desire a consistent holographic realization of our embedding, since we ultimately want to share the argument of (non-) integrability with the dual field theory as well. Hence, one should care about the validity of our embedding (and of every other embedding for that matter) on this kind of space. A first answer has already been provided through \cite{Dobashi:2002ar}, where the string trajectories on the geometry (\ref{PoincareOmega}) are shown to naturally realize the holographic principle. The second argument has to do with our particular formulation. The dual field theory lives on $\mathbb{R}^4$, in which the target space time of our interest lives, i.e. $t\equiv x_1$. Since we have re-expressed $\mathbb{R}^4$ in the spherical coordinates (\ref{CartToSphere}), then the radial coordinate $r$ should incorporate (Euclidean) time. Therefore, since we do include $r(\tau)$ into our dynamics, through the equation of motion (\ref{EOMr}), everything is in order and our string should have a well-defined holographic realization.\\

\paragraph{\textbf{A simple solution}}
Next, we desire a simple solution on the equations of motion, around which we can study the fluctuations of the string. In that respect, regardless of having used the symmetries of the background to simplify our embedding (towards a less symmetric truncation), the equations of motion (\ref{EOMr})-(\ref{EOMphi}) still possess a rich variety of simple solutions. However, not all of these solutions are consistent with our particular embedding: any consistent solution must also satisfy the worldsheet constraint (\ref{VC}). Given, in turn, the set of the consistent simple solutions, not all of those are actually useful since not all of them permit fluctuations that include the $B_2$ field contribution on the dynamics. The latter being the only possible non-integrable deviation from the integrable H$_5\times$dS$_5$. The associated $H_3$ flux dynamics is reflected on the $\beta$-dependent terms in (\ref{EOMr})-(\ref{EOMphi}), thus our simple solution should let those terms survive in our fluctuating equations.\vspace{0.15cm}

Under the above considerations, it turns out that there is an infinite set of invariant planes that do the job, for $\theta\in(\frac{\pi}{4},\frac{\pi}{2})\cup(\frac{\pi}{2},\frac{3\pi}{4})$ and $\chi\in(0,\pi)$. It may seem naively odd, but the most $-$ by far $-$ convenient choice comes with the invariant plane
\begin{equation}
\begin{split}
\Big\lbrace r=\dot{r}=\ddot{r}=0,\;&\chi=\frac{\pi}{2},\,\dot{\chi}=\ddot{\chi}=0,\\
&\theta=\theta_\star\equiv\arctan\sqrt{\frac{5}{3}},\,\dot{\theta}=\ddot{\theta}=0\Big\rbrace\label{Plane}
\end{split}
\end{equation}
around which the fluctuations simplify tremendously. On this plane, the equations of motion (\ref{EOMr})-(\ref{EOMphi}) are satisfied along with the simple solutions
\begin{equation}
\begin{split}
\phi(\tau)\;&=\;-\nu\tau\\
z(\tau)\;&=\;\frac{e^{\frac{\nu}{2}\tau}}{\sqrt{10}}\label{phiSimpleSol}
\end{split}
\end{equation}
where the coefficients including the winding number $\nu$ were identified by the Virasoro constraint (\ref{VC}), while the signs and the constants were selected to our convenience without loss of generality \footnote{To be precise, there is another choice of signs that gives a similar result, while the rest of the choices turn out quite complicated.}.\vspace{0.15cm}

Note that the symmetry analysis on the background was not necessary to build an embedding. We just used it to shape a less symmetric string truncation, so as to have a better chance in non-integrability. Had we not used those symmetry considerations, we would have chosen a far more general embedding whose equations of motion would include a large variety of invariant planes. Nevertheless, all of those planes would eventually descend down to the invariant plane (\ref{Plane}) and its corresponding simple solution (\ref{phiSimpleSol}) as the \textit{only} useful option, just through a way more laborious path.\vspace{0.15cm}

Next, we expand around the invariant plane in order to study the dynamical behavior of the system there. While the $r, \chi$ and $\theta$ fluctuations around the plane are generally coupled, such complexity is not eventually needed in our case. Stated otherwise, we shall study isolated fluctuations on each one of those dimensions, around the invariant plane (\ref{Plane}) and on the simple solution (\ref{phiSimpleSol}). We call such a fluctuation a Normal Variational Equation (NVE).\\

\paragraph{\textbf{Fluctuations around the invariant plane}}
To isolate the $\theta$-fluctuations around the invariant plane (\ref{Plane}), we expand as $\theta(\tau)=\theta_\star+\epsilon\,\vartheta(\tau)$ for $\epsilon\rightarrow0$ in the $\theta$-equation of motion (\ref{EOMtheta}), while we keep the other dimensions frozen, i.e. $\lbrace r=\dot{r}=\ddot{r}=0,\;\chi=\frac{\pi}{2},\,\dot{\chi}=\ddot{\chi}=0\rbrace$. Hence, we obtain the $\theta$-NVE
\begin{equation}
\ddot{\vartheta}(\tau)\;=\;0
\end{equation}
which has a Liouvillian solution.\vspace{0.15cm}

In the same vein, the isolated $\chi$-fluctuations around the invariant plane occur for $\chi(\tau)=\frac{\pi}{2}+\epsilon\,x(\tau)$ while $\lbrace r=\dot{r}=\ddot{r}=0,\:\theta=\theta_\star,\,\dot{\theta}=\ddot{\theta}=0\rbrace$, which however solves the $\chi$-equation of motion (\ref{EOMchi}) identically and gives no further insight.\vspace{0.15cm}

Therefore, we are only left with the $r$-fluctuations around the invariant plane (\ref{Plane}). To isolate those, we expand as $r(\tau)=0+\epsilon\,\varrho(\tau)$ for $\epsilon\rightarrow0$ in the $r$-equation of motion (\ref{EOMr}), while we keep the other dimensions frozen, i.e. $\lbrace\chi=\frac{\pi}{2},\,\dot{\chi}=\ddot{\chi}=0,\:\theta=\theta_\star,\,\dot{\theta}=\ddot{\theta}=0\rbrace$. Hence, we obtain the $r$-NVE
\begin{equation}
\ddot{\varrho}(\tau)\:-\:\nu\dot{\varrho}(\tau)\:+\:\kappa\left(\kappa+\frac{\beta\nu\,e^{\frac{3\nu}{2}\tau}}{32}\right)\varrho(\tau)\;=\;0\label{rNVE}
\end{equation}
which is solved for
\begin{equation}
\begin{split}
\varrho(\tau)\:&=\:c_1\,\mathrm{J}_{G}\left(f(\tau)\right)\,e^{\frac{\nu\tau}{2}}\,\Gamma\left(1+G\right)\hspace{2cm}\\
&\hspace{3cm}+c_2\,\mathrm{J}_{-G}\left(f(\tau)\right)\,e^{\frac{\nu\tau}{2}}\,\Gamma\left(1-G\right)\\[10pt]
f(\tau)\:&=\:\sqrt{\frac{\beta\kappa\,e^{\frac{3\nu\tau}{2}}}{18\nu}}\;,\hspace{0.7cm}G\:=\:\frac{2\sqrt{-4\kappa^2+\nu^2}}{3\nu}
\end{split}\label{rsolution}
\end{equation}
where $c_1,c_2$ are constants and $\mathrm{J}_n(\tau), \Gamma(z)$ are the Bessel function of the first kind and the gamma function, respectively. Before anything, two comments are in place here. First, if the string windings are such that $\kappa\,\nu<0$, then $f(\tau)\in\mathbb{I}$ and we just work with the modified Bessel functions. Secondly, if the windings are such that $G\in\mathbb{I}$ then $\mathrm{J}_n(\tau)$ acquires a purely imaginary order $n\in\mathbb{I}$ and gives a complex number $z_1(\tau)\in\mathbb{C}$, while its conjugate function $\mathrm{J}_{-n}(\tau)$ gives $z_1^\star(\tau)$. Similarly, $\Gamma(z)$ with $z\in\mathbb{C}$ gives a complex number $z_2\in\mathbb{C}$, while $\Gamma(z^\star)$ gives $z_2^\star$. Thus, for $G\in\mathbb{I}$, our $\varrho$-solution (\ref{rsolution}) can be written as\vspace{0.15cm}
\begin{equation}
\varrho(\tau)\;=\;c_1\,e^{\frac{\nu\tau}{2}}\,z_1(\tau)z_2\;+\;c_2\,e^{\frac{\nu\tau}{2}}\,z_1^\star(\tau)z_2^\star
\end{equation}
which can only be real for $c_1=c_2$. This is a necessary condition for the physicality of our solution.\vspace{0.15cm}

The Bessel function is non-Liouvillian except only for half integer order $n$. If $n=\pm G$ is imaginary then it can never be a half integer, anyway. If it is real, on the other hand, $\pm G$ reflects the various configurations of our embedding and thus it cannot be restricted without losing generality. In other words, we should care about the solution (\ref{rsolution}) on every value of the winding numbers $\kappa,\nu$. Even if there are particular string configurations (for appropriate $\kappa,\nu$) that are Liouvillian, there are always others that are not. Hence, we have ultimately spotted a string embedding that exhibits non-integrable dynamics.\vspace{0.15cm}

As a consistency check, note that for $\beta=0$ in (\ref{rNVE}) we recover integrability, as we should for an undeformed and symmetric vacuum. The same holds for $\kappa,\nu=0$, where the string reduces to a point particle on H$_5\times$dS$_5$ that cannot feel the $H_3$ flux.\\

\paragraph{\textbf{A simpler embedding}} Since one is never enough, we shall study another string embedding. We have already mentioned that had we included extra string coordinate dependence than the one we chose before, we would have ultimately ended up studying the embedding (\ref{stringSoliton}). Hence, we are lead to build a simpler truncation this time. It turns out that our most minimal alternative is to localize the coordinates $z=z_0=1$ and $\chi=\frac{\pi}{2}$ in our previous embedding, i.e.
\begin{equation}
\begin{split}
r=r(\tau)\hspace{0.6cm}\chi=\frac{\pi}{2}\hspace{0.6cm}\psi=\frac{\pi}{2}\hspace{0.6cm}\xi=\kappa\sigma\hspace{0.6cm}z=1\hspace{0.7cm}\\[10pt]
\theta=\theta(\tau)\hspace{0.7cm}\phi=\phi(\tau)\hspace{0.7cm}\omega_1=\omega_2=\frac{\pi}{2}\hspace{0.7cm}\omega_3=\nu\sigma\hspace{0.5cm}
\end{split}\label{stringSoliton2}
\end{equation}
where $\kappa,\nu\in\mathbb{Z}$. The $B_2$ field that couples to the new embedding reduces to
\begin{equation}
B_2=-\frac{\beta}{4}re^{-\phi}\sin\theta\d r\wedge\d\xi
\end{equation}
while the associated Lagrangian density becomes
\begin{equation}
\begin{split}
\mathcal{L}\;=\;-\dot{r}^2+\kappa^2r^2-\dot{\theta}^2+\sin^2\theta\:&\dot{\phi}^2+\nu^2\cos^2\theta\\
&-\frac{\beta\kappa}{2}re^{-\phi}\sin\theta\:\dot{r}
\end{split}\label{Lagrangian2}
\end{equation}
Of course, on this embedding too, the equations of motion for this Lagrangian are equivalent to those of the $\sigma$-model and read
\begin{equation}
4\ddot{r}\;=\;-4\kappa^2r\:+\:\beta\kappa\,re^{-\phi}\left(\sin\theta\:\dot{\phi}-\cos\theta\:\dot{\theta}\right)\label{EOMr2}
\end{equation}\vspace{-15pt}
\begin{equation}
\begin{split}
4\ddot{\theta}\;=\;2\sin2\theta\left(\nu^2-\dot{\phi}^2\right)\:+\:\beta\kappa\,e^{-\phi}\cos\theta\,r\:\dot{r}\label{EOMtheta2}
\end{split}
\end{equation}\vspace{-10pt}
\begin{equation}
4\sin\theta\:\ddot{\phi}\;=\;-8\cos\theta\,\:\dot{\theta}\dot{\phi}\:+\:\beta\kappa\,e^{-\phi}\,r\:\dot{r}\hspace{1cm}\label{EOMphi2}
\end{equation}
These equations are constrained by the worldsheet equation of motion, i.e. the Virasoro constraint
\begin{equation}
\begin{split}
2\,T_{\tau\tau}&=\:2\,T_{\sigma\sigma}\,=\,\dot{r}^2+\kappa^2r^2-\sin^2\theta\:\dot{\phi}^2+\dot{\theta}^2+\nu^2\cos^2\theta=0\\[5pt]
T_{\tau\sigma}&=0
\end{split}\label{VC2}
\end{equation}
The worldsheet energy-momentum tensor is conserved, $\nabla_aT^{ab}=0$, since $\partial_\tau T_{\tau\tau}=\partial_\sigma T_{\sigma\sigma}=0$ on the equations of motion (\ref{EOMr2})-(\ref{EOMphi2}), yielding also the consistency of our embedding. Of course, the associated Hamiltonian system here is qualitatively the same as with the previous embedding.\vspace{0.15cm}

In this particular case however, under the considerations $-$ again $-$ of consistency and of including the $H_3$ flux contribution, there is only one invariant plane that serves our cause. That is
\begin{equation}
\Big\lbrace r=\dot{r}=\ddot{r}=0,\;\theta=\frac{\pi}{4},\,\dot{\theta}=\ddot{\theta}=0\Big\rbrace\label{Plane2}
\end{equation}
Note that, while for the previous embedding the choice $\theta=\frac{\pi}{4}$ was excluded since it lead to useless invariant planes, here it constitutes our only option. This is indeed the \textit{unique} plane that does the job and on which the equations of motion (\ref{EOMr2})-(\ref{EOMphi2}) are satisfied, along with the simple solution
\begin{equation}
\phi(\tau)\;=\;-\nu\tau\label{simplesol2}
\end{equation}
where the coefficient was identified with the winding number $\nu$ through the Virasoro constraint (\ref{VC2}), while the sign was again selected for our convenience without loss of generality. For one last time, we move on to study the isolated fluctuations around the invariant plane (\ref{Plane2}) and on its associated simple solution (\ref{simplesol2})\vspace{0.15cm}

Obviously, the $\theta$-fluctuations are again trivial and so we are left to study the fluctuations along $r$. We expand $r(\tau)=0+\epsilon\,\varrho(\tau)$ for $\epsilon\rightarrow0$ in the $r$-equation of motion (\ref{EOMr2}), while we keep $\theta$ frozen, i.e. $\lbrace\theta=\frac{\pi}{4},\dot{\theta}=\ddot{\theta}=0\rbrace$. Hence, we obtain the $r$-NVE
\begin{equation}
\ddot{\varrho}(\tau)\:+\:\kappa\left(\kappa+\frac{\sqrt{2}\,\beta\nu\,e^{\nu\tau}}{8}\right)\varrho(\tau)\;=\;0\label{rNVE2}
\end{equation}
which is solved for
\begin{equation}
\begin{split}
\varrho(\tau)\:&=\:c_1\,\mathrm{J}_{G}\left(f(\tau)\right)\,\Gamma\left(1+G\right)\hspace{2cm}\\
&\hspace{3cm}+c_2\,\mathrm{J}_{-G}\left(f(\tau)\right)\,\Gamma\left(1-G\right)\\[10pt]
f(\tau)\:&=\:\sqrt{\frac{\beta\kappa\,e^{\nu\tau}}{\sqrt{2}\nu}}\;,\hspace{0.7cm}G\:=\:\frac{2i\kappa}{\nu}
\end{split}\label{rsolution2}
\end{equation}
where $c_1,c_2$ are constants and $\mathrm{J}_n(\tau), \Gamma(z)$ are the Bessel function of the first kind and the gamma function, respectively. Again, if the string windings are such that $\kappa\,\nu<0$, then $f(\tau)\in\mathbb{I}$ and we just work with the modified Bessel functions. Also, as explained for the case of the previous solution (\ref{rsolution}), since the order $n=\pm G$ of the Bessel function is purely imaginary, it can never be a half integer (that gives a Liouvillian solution) while it must necessarily hold that $c_1=c_2$ for the physicality of our solution (\ref{rsolution2}). Hence, we have spotted another non-integrable fluctuation of the string.\vspace{0.15cm}

Again, as a consistency check, note that for $\beta=0$ in (\ref{rNVE2}) we recover integrability, as we should for the undeformed vacuum. The same holds for $\kappa,\nu=0$, where the string reduces to a point particle on H$_5\times$dS$_5$ that does not couple to the Kalb-Ramond field.\vspace{0.15cm}

As indicated repeatedly, the invariant planes we have studied so far are the unique solutions that consistently incorporate the $H_3$ flux contribution. Nevertheless, in case we want to be persistent and make the non-integrable character of the system manifest in an additional way, we could go for a more involved string embedding. In particular, we could build a \textit{spinning string} by letting
\begin{equation}
\xi(\tau,\sigma)=\kappa\sigma+\Xi(\tau)\hspace{1cm}\omega_3(\tau,\sigma)=\nu\sigma+\Omega(\tau)
\end{equation}
in the previous embeddings, (\ref{stringSoliton}) and (\ref{stringSoliton2}). Choosing that truncation, worldsheet consistency conditions (on necessarily similar invariant planes) drop the dynamics down to the exact same results we found for the simpler embeddings.\vspace{0.15cm}

As an additional consistency check, we can repeat everything we have done so far in Euclidean signature, i.e. on the Euclidean AdS$_5\times$S$^5$. In order to do this, we Wick rotate the target space in (\ref{PoincareOmega}) as $\phi\rightarrow i\varphi$ while we pick $-$ for consistency $-$ an also Euclidean worldsheet. Again, we acquire the exact same results up to certain factors.\vspace{0.15cm}

Since an integrable structure exhibits its homonymous property on all of its sectors, we deduce that the dynamical sector we studied and, therefore, the whole supergravity background under consideration are classically non-integrable.

\section{Epilogue}
Ultimately, we have proven that string theory on the vacuum dual to the $\Omega$-deformed $\mathcal{N}=4$ SYM, recently proposed in \cite{Bobev:2019ylk}, is classically non-integrable in the Liouvillian sense. Using the broken symmetries (by the $H_3$ flux) of the background, we constructed appropriate string embeddings and studied their fluctuations around simple solutions of their equations of motion. Since particular fluctuations turned out to be non-Liouvillian for a general string configuration, we declared the whole theory as non-integrable.\vspace{0.15cm}

Notice that, contrary to the usual method of analytic non-integrability, on this particular analysis we did not have to enforce differential Galois theory and Kovacic's theorem on differential equations of motion. That is, we reached exact non-Liouvillian solutions given in terms of the Bessel function of the first kind, of no half-integer order for a general string configuration.\vspace{0.15cm}

Since the supergravity background we examined is dual to the $\Omega$-deformed $\mathcal{N}=4$ SYM, holography dictates that the statement of non-integrability must be shared by the gauge theory as well. Hence, apart from being just a delicate Hamiltonian mechanics problem, the present work suggests that the $\Omega$-deformation does not preserve classical integrability.\vspace{0.15cm}

However, a non-integrable theory may possess integrable subsectors or limits. In the $\Omega$-deformed theory, this is obviously true on the grounds of the existing literature that associates this deformation with various integrable structures, as noted in the introduction. Therefore, the ontology of the regimes of integrability is worthy of further examination. More interestingly though, given the $\Omega$ dual background (\ref{PoincareOmega}), a valuable study would be based on its Kalb-Ramond field which realizes the $\Omega$-deformation itself. In particular, special vacua or limits of string theory on this supergravity background could investigate the action of this $B_2$ field on the associated string states, while $-$ in that case $-$ holography should be in place to shed light on their dual $\Omega$-deformed field theory subsectors.\\

\paragraph{\textbf{Acknowledgments}} I thank D. Thompson, C. Nunez, N. Bobev and D. Giataganas for their comments. This work was supported by a STFC scholarship. It is dedicated to the memory of Alex Grigoropoulos.

\appendix


\bibliographystyle{apsrev4-1} 

\end{document}